\documentstyle[11pt,moriond,epsfig]{article}

\def\be{\begin{equation}}
\def\ee{\end{equation}}
\def\bea{\begin{eqnarray}}
\def\eea{\end{eqnarray}}

\begin{document}
\vspace*{4cm}
\title{RESULTS ON CASCADE PRODUCTION IN LEAD-LEAD INTERACTIONS
\\ FROM THE NA57 EXPERIMENT}

\author{Presented by D. Elia for the NA57 Collaboration:
\\
\vspace{2mm}
F.~Antinori$^{k}$,
A.~Badal{\`a}$^{f}$,
R.~Barbera$^{f}$,
A.~Bhasin$^{d}$,
I.J.~Bloodworth$^{d}$,
G.E.~Bruno$^{a}$,
S.A.~Bull$^{d}$,
R.~Caliandro$^{a}$,
M.~Campbell$^{g}$,
W.~Carena$^{g}$,
N.~Carrer$^{g}$,
R.F.~Clarke$^{d}$,
A.P.~de~Haas$^{r}$,
P.C.~de~Rijke$^{r}$,
D.~Di~Bari$^{a}$,
S.~Di~Liberto$^{n}$,
R.~Divia$^{g}$,
D.~Elia$^{a}$,
D.~Evans$^{d}$,
K.~Fanebust$^{b}$,
F.~Fayazzadeh$^{j}$,
J.~Fedorisin$^{i}$,
G.A.~Feofilov$^{p}$,
R.A.~Fini$^{a}$,
J.~Ft\'a\v cnik$^{e}$,
B.~Ghidini$^{a}$,
G.~Grella$^{o}$,
H.~Helstrup$^{c}$,
M.~Henriquez$^{j}$,
A.K.~Holme$^{j}$,
A.~Jacholkowski$^{a}$,
G.T.~Jones$^{d}$,
P.~Jovanovic$^{d}$,
A.~Jusko$^{h}$,
R.~Kamermans$^{r}$,
J.B.~Kinson$^{d}$,
K.~Knudson$^{g}$,
A.A.~Kolozhvari$^{p}$,
V.~Kondratiev$^{p}$,
I.~Kr\'alik$^{h}$,
A.~Kravcakova$^{i}$,
P.~Kuijer$^{r}$,
V.~Lenti$^{a}$,
R.~Lietava$^{e}$,
G.~L\o vh\o iden$^{j}$,
M.~Lupt\'ak$^{h}$,
V.~Manzari$^{a}$,
G.~Martinska$^{i}$,
M.A.~Mazzoni$^{n}$,
F.~Meddi$^{n}$,
A.~Michalon$^{q}$,
M.~Morando$^{k}$,
D.~Muigg$^{r}$,
E.~Nappi$^{a}$,
F.~Navach$^{a}$,
P.I.~Norman$^{d}$,
A.~Palmeri$^{f}$,
G.S.~Pappalardo$^{f}$,
B.~Pastir\v c\'ak$^{h}$,
J.~Pisut$^{e}$,
N.~Pisutova$^{e}$,
F.~Posa$^{a}$,
E.~Quercigh$^{k}$,
F.~Riggi$^{f}$,
D.~R\"ohrich$^{b}$,
G.~Romano$^{o}$,
K.~\v{S}afa\v{r}\'{i}k$^{g}$,
L.~\v S\'andor$^{h}$,
E.~Schillings$^{r}$,
G.~Segato$^{k}$,
M.~Sen\`e$^{l}$,
R.~Sen\`e$^{l}$,
W.~Snoeys$^{g}$,
F.~Soramel$^{k}$,
P.~Staroba$^{m}$,
T.A.~Toulina$^{p}$,
R.~Turrisi$^{k}$,
T.S.~Tveter$^{j}$,
J.~Urb\'{a}n$^{i}$,
F.~Valiev$^{p}$,
A.~van~den~Brink$^{r}$,
P.~van~de~Ven$^{r}$,
P. Vande Vyvre$^{g}$,
N.~van~Eijndhoven$^{r}$,
J.~van~Hunen$^{g}$,
A.~Vascotto$^{g}$,
T.~Vik$^{j}$,
O.~Villalobos Baillie$^{d}$,
L.~Vinogradov$^{p}$,
T.~Virgili$^{o}$,
M.F.~Votruba$^{d}$,
J.~Vrl\'{a}kov\'{a}$^{i}$ and
P.~Z\'{a}vada$^{m}$
\vspace{2mm}
}
\address{$^{a}$ Dipartimento IA di Fisica dell'Universit{\`a}
       e del Politecnico
       di Bari and INFN, Bari, Italy \\
$^{b}$ Fysisk Institutt, Universitetet i Bergen, Bergen, Norway\\
$^{c}$ H{\o}gskolen i Bergen, Bergen, Norway\\
$^{d}$ University of Birmingham, Birmingham, UK\\
$^{e}$ Comenius University, Bratislava, Slovakia\\
$^{f}$ University of Catania and INFN, Catania, Italy\\
$^{g}$ CERN, European Laboratory for Particle Physics, Geneva,
       Switzerland\\
$^{h}$ Institute of Experimental Physics, Slovak Academy of Science,
       Ko\v{s}ice, Slovakia\\
$^{i}$ P.J. \v{S}af\'{a}rik University, Ko\v{s}ice, Slovakia\\
$^{j}$ Fysisk Institutt, Universitetet i Oslo, Oslo, Norway\\
$^{k}$ University of Padua and INFN, Padua, Italy\\
$^{l}$ Coll\`ege de France, Paris, France\\
$^{m}$ Institute of Physics, Prague, Czech Republic\\
$^{n}$ University ``La Sapienza'' and INFN, Rome, Italy\\
$^{o}$ Dipartimento di Scienze Fisiche ``E.R. Caianiello''
       dell'Universit{\`a} and INFN, Salerno, Italy\\
$^{p}$ State University of St. Petersburg, St. Petersburg, Russia\\
$^{q}$ Institut de Recherches Subatomique, IN2P3/ULP, Strasbourg, France\\
$^{r}$ Utrecht University and NIKHEF, Utrecht, The Netherlands
}

\maketitle

\abstracts{
The NA57 experiment has been designed to
study the production of strange and multi-strange
particles in
Pb-Pb and p-Be collisions at the CERN SPS.
\\
The predecessor experiment WA97 has measured
an enhanced abundance of strange particles in Pb-Pb
collisions relative to p-A reactions
at 160 GeV/$c$ per nucleon beam momentum.
NA57 has extended the WA97 measurements
to investigate
the evolution of the strangeness enhancement pattern as a function of
the beam energy and over a wider centrality range.
In this paper, we report
results on $\Xi^-$ and $\overline\Xi^+$
hyperon production
for about the 60\% most central Pb-Pb collisions at
160 GeV/$c$ per nucleon.
}

\setcounter{figure}{0}

\section{Introduction}

The WA97 experiment has measured an enhancement
in the production of strange and multi-strange baryons and anti-baryons
when going from p-Be to central
Pb-Pb collisions~\cite{Ant99}.
The observed effect increases with the strangeness content
of the particles. Such a behaviour 
has been predicted~\cite{RafMul82} as a consequence of the QCD
phase transition to a Quark Gluon Plasma (QGP) and
is not reproduced by
microscopic hadronic collision models.
\\
The main goal of NA57 is to study the dependence 
of the enhancement on the interaction volume and on the
collision energy per incoming nucleon~\cite{NA57prop}. 
To this purpose, 
the experiment has extended the centrality range down to
a lower limit of about 50 wounded nucleons (the corresponding
limit for WA97 was about 100) and has collected data using
both 160 and 40 $A$ GeV/$c$ beams at the CERN SPS.

\section{Experimental apparatus and data sets}

The NA57 experiment detects strange and multi-strange
hyperons by reconstructing their weak decays
into final states with charged particles only
(e.g. $\Xi^-$ $\rightarrow$ $\Lambda\pi^-$, with 
$\Lambda$ $\rightarrow$ $\pi^-{p}$).
Tracks are measured in the silicon telescope, an
array of 13 pixel detector planes with 5 $\times$ 5 cm$^2$
cross section and 30 cm length, placed 60 cm downstream of the target
and inclined with respect to the beam line to
match the central rapidity region. 
Similarly to WA97, the
full acceptance coverage corresponds to about one unit of 
rapidity at medium transverse momentum.
The centrality trigger for Pb-Pb interactions, based on a scintillator
petal system placed 10 cm downstream of the target,
selects the most central 60\% of the inelastic
cross section. The centrality of the collision
is determined from the charged particle multiplicity
sampled at central rapidity by two silicon strip detector
stations (MSD).
A detailed description of the full apparatus can be found 
elsewhere~\cite{Man99}.
\\
A summary of the NA57 data samples is reported
in Table~\ref{tab:NA57data}.
The experiment has collected data on Pb-Pb collisions
at both 160 and 40 $A$ GeV/$c$ beam momentum. 
A small reference data sample of p-Be interactions 
at 40 GeV/$c$ was collected 
in 1999. The collection of p-Be reference data
will continue in the summer 2001.

\begin{table}[h]
\caption{Data sets and status 
of reconstruction.\label{tab:NA57data}} 
\vspace{0.4cm}
\begin{center}
\begin{tabular}{|r|r|r|r|r|}
\hline
{\bf System} & {\bf Beam mom.} &
{\bf Sample size} & {\bf Data taking} 
& {\bf Reconstruction} \\ \hline
{Pb Pb} &
{160 A GeV/c} & {230 M events} &
{November 1998} & {February 2000} \\
{p Be} &
{40 GeV/c} & {60 M events} &
{July 1999} & {July 2000} \\
{Pb Pb} &
{40 A GeV/c} & {290 M events} &
{November 1999} & {January 2001} \\
{Pb Pb} &
{160 A GeV/c} & {230 M events} &
{October 2000} & {October 2001 (exp)} \\
{p Be} &
{40 GeV/c} & {150 M events (exp)} &
{August 2001} & {December 2001 (exp)} \\ \hline
\end{tabular}
\end{center}
\end{table}

\noindent Reference data on p-Be and p-Pb at 160 GeV/$c$ 
are available from the WA97 measurements.
The results on
$\Xi^-$ and $\overline\Xi^+$ particles
reported in this paper 
have been obtained
analyzing the 1998 Pb-Pb data sample at
160 $A$ GeV/$c$.

\section{Data analysis and results}

\begin{figure}[h]
\centering
  \includegraphics[clip,scale=0.4]{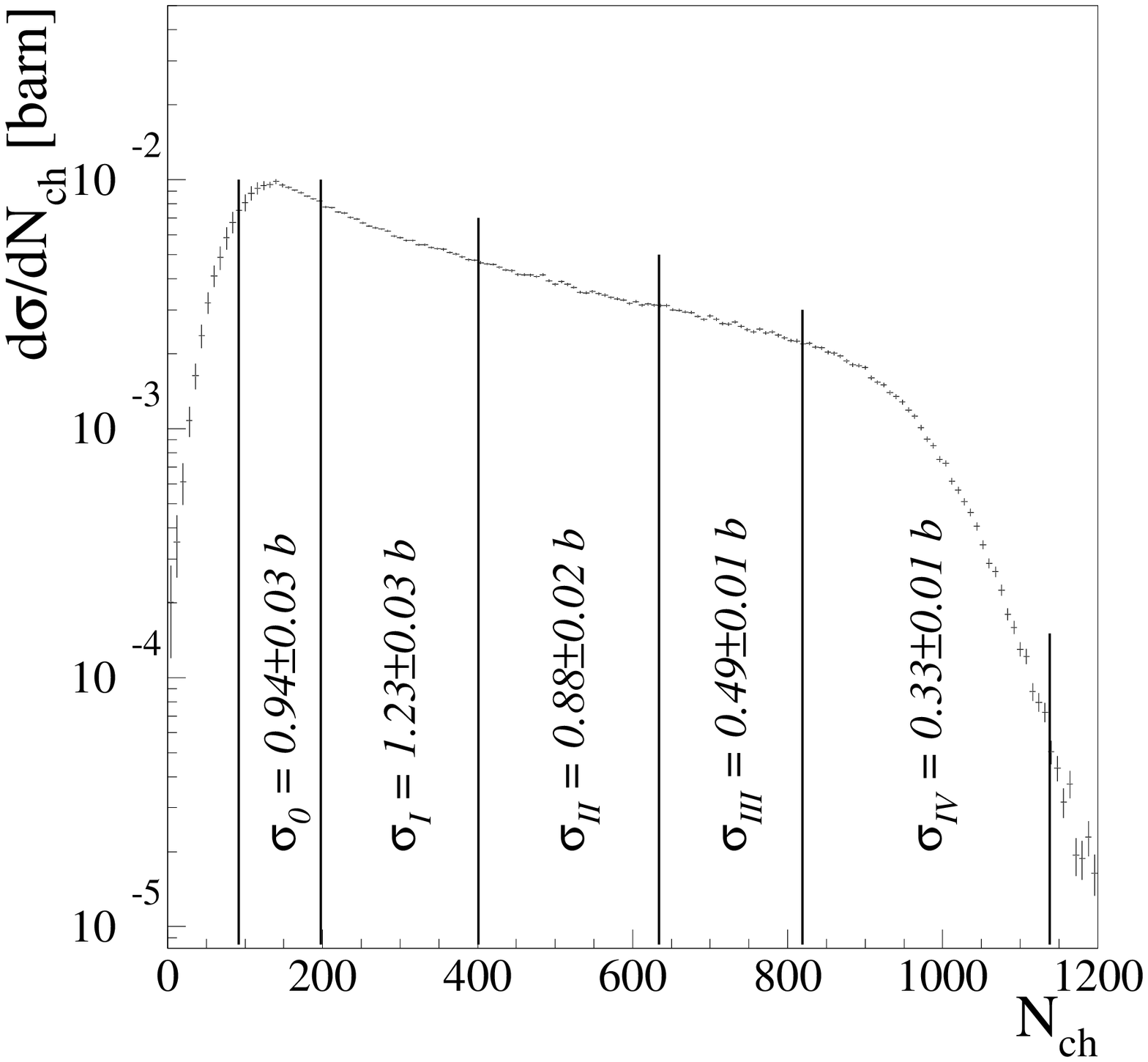}
    \hspace{0.3cm}
   \includegraphics[clip,scale=0.4]{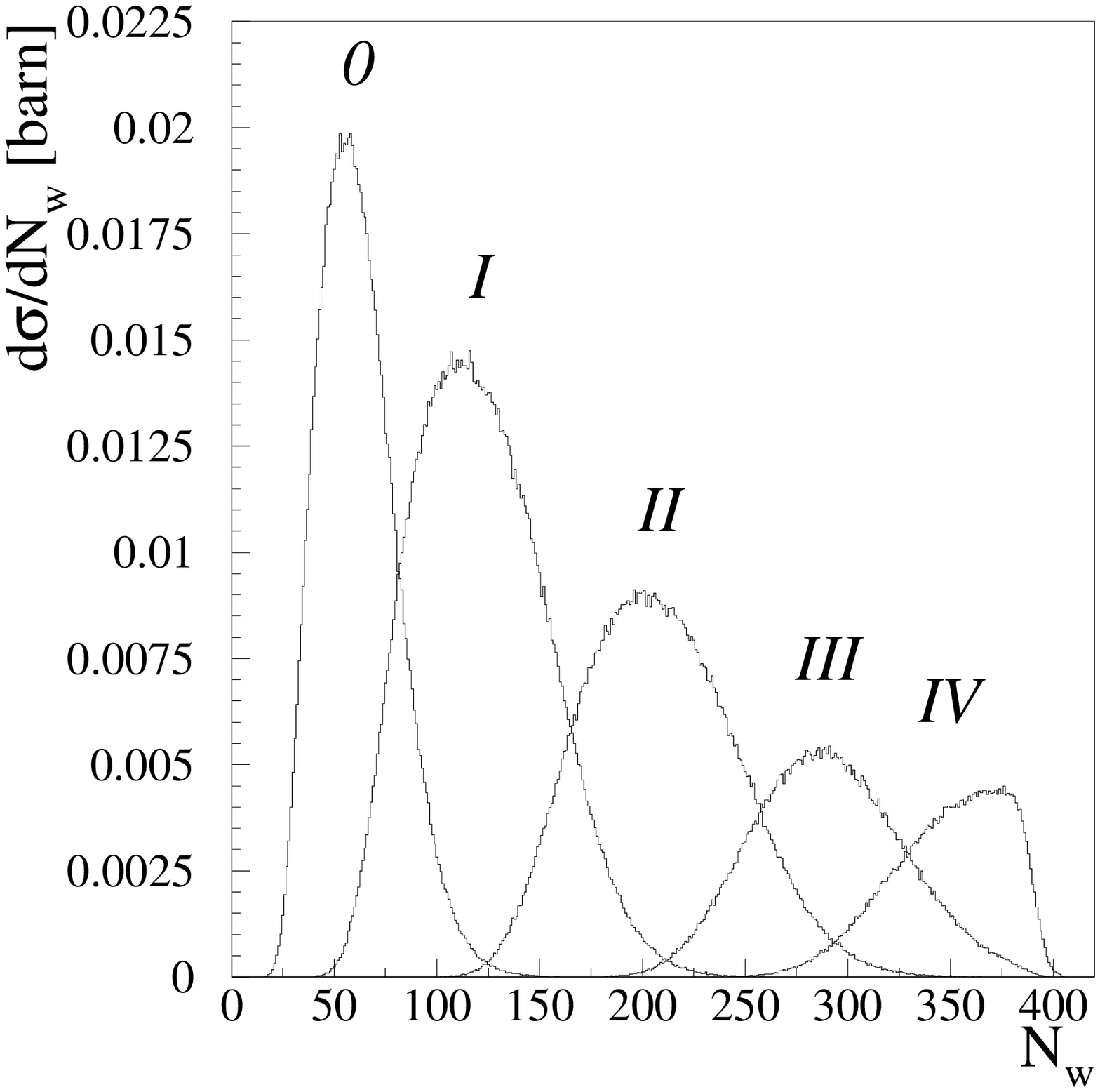}
\caption{Multiplicity distribution (left) and
distributions of wounded nucleons in each multiplicity
class (right).\label{fig:mult}} 
\end{figure}

The collision centrality has been evaluated
by measuring the charged particle multiplicity
in the interval 2 $<$ $\eta$ $<$ 4
by the MSD.
The multiplicity spectrum, divided into
five bins, is shown in the
left hand plot on Fig.\ref{fig:mult}:
the drop at low multiplicities 
is due to the
centrality trigger condition.
The measured cross section corresponding to each
class is also indicated.
The distributions of the number of wounded nucleons,
obtained in the framework of the Glauber Model, are
shown on the right hand plot~\cite{carrers2000}.
Classes I to IV correspond
to the four centrality bins of
WA97.

\begin{figure}[h]
\centering
  \includegraphics[clip,scale=0.38]{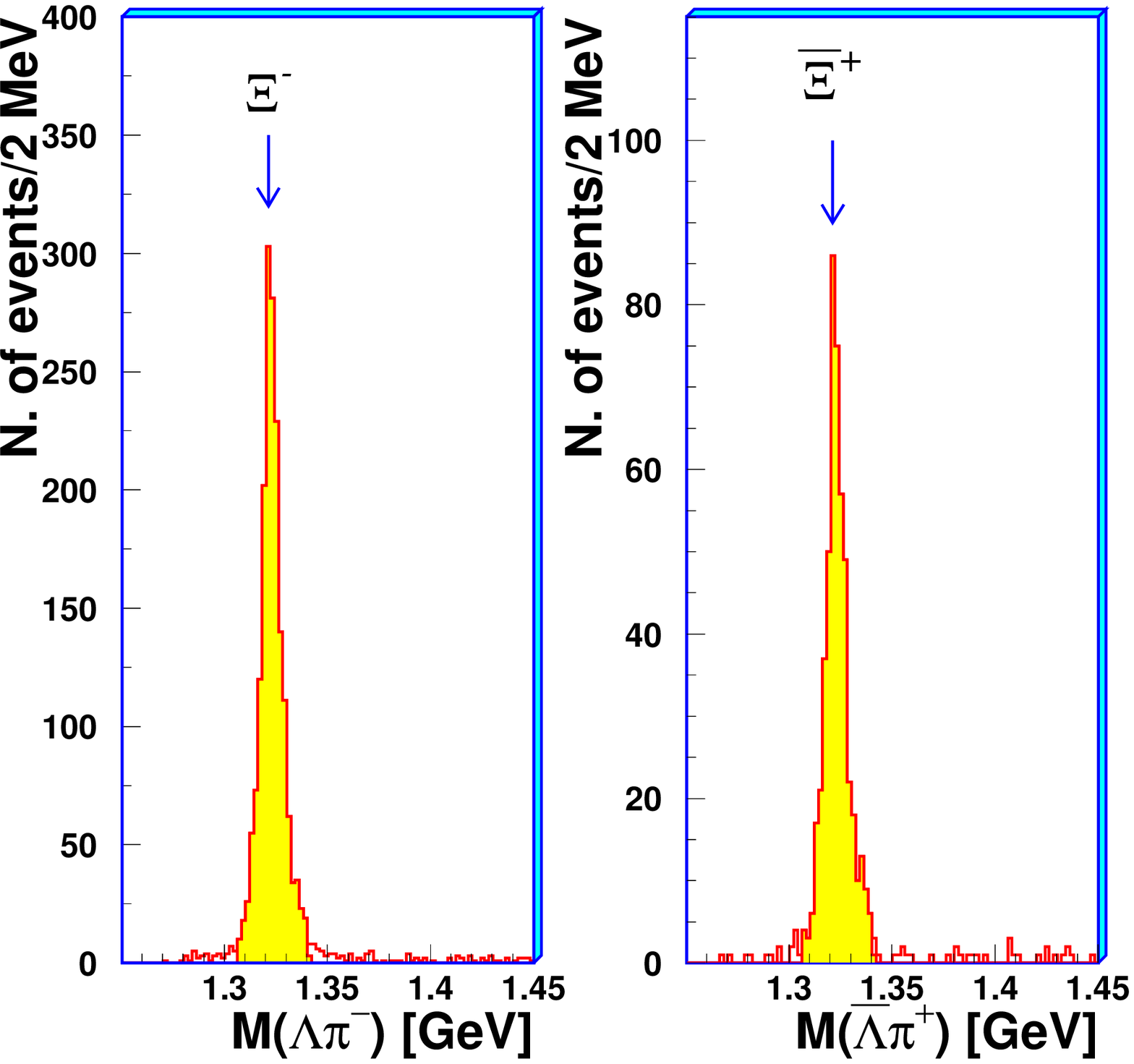}
    \hspace{0.3cm}
   \includegraphics[clip,scale=0.38]{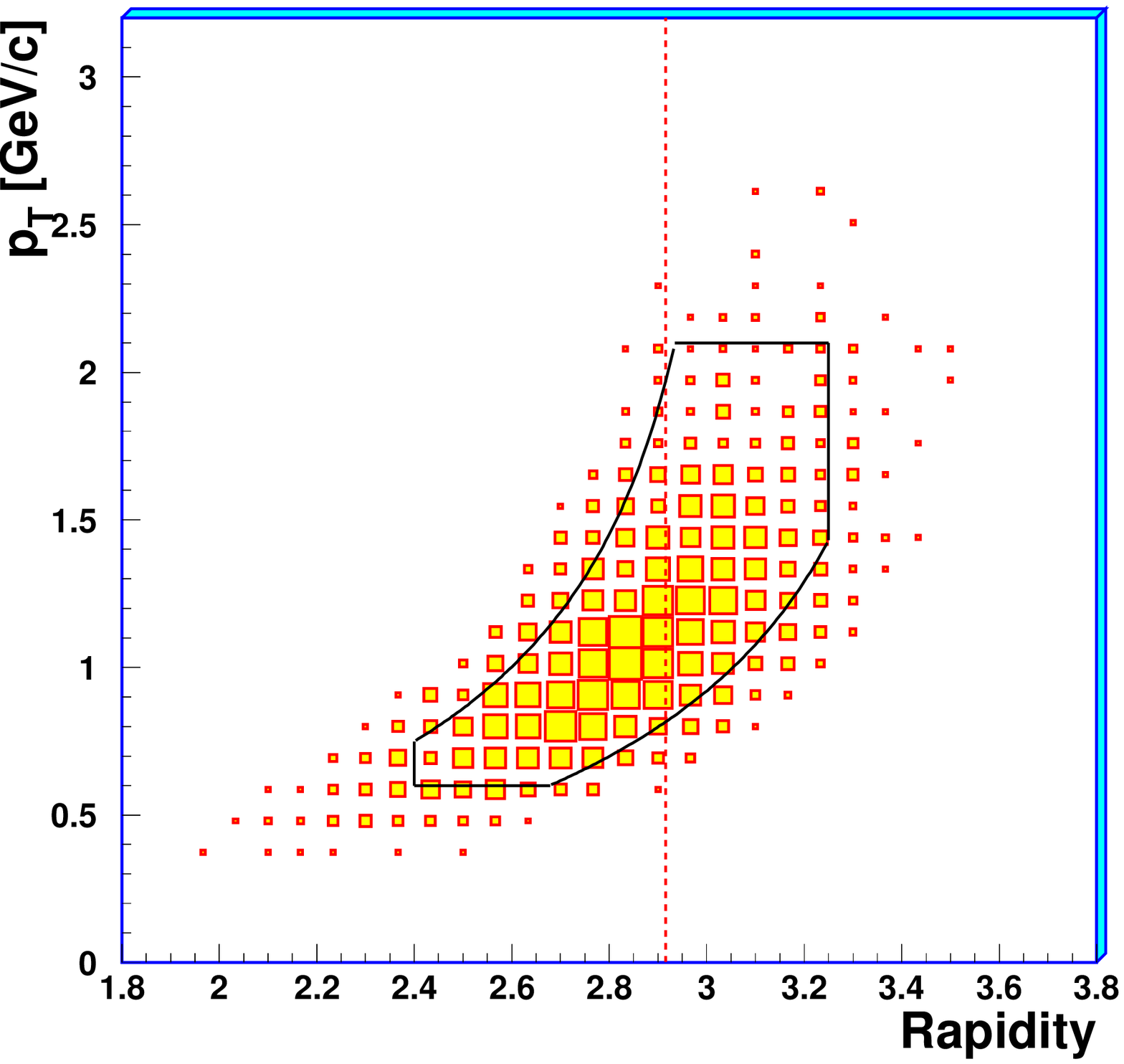}
\caption{$\Lambda\pi$ invariant mass distributions (left) and
corresponding p$_T$ versus rapidity distribution (right).\label{fig:cassig}}
\end{figure}

\noindent 
Fig.\ref{fig:cassig} shows the $\Lambda\pi$ invariant mass
distributions after all the selection cuts and the 
corresponding kinematic window (enclosed area) selected 
in the y-p$_T$ distribution of the reconstructed 
$\Xi^-$ and $\overline\Xi^+$ particles.

\begin{figure}[h]
\centering
  \includegraphics[clip,scale=0.48]{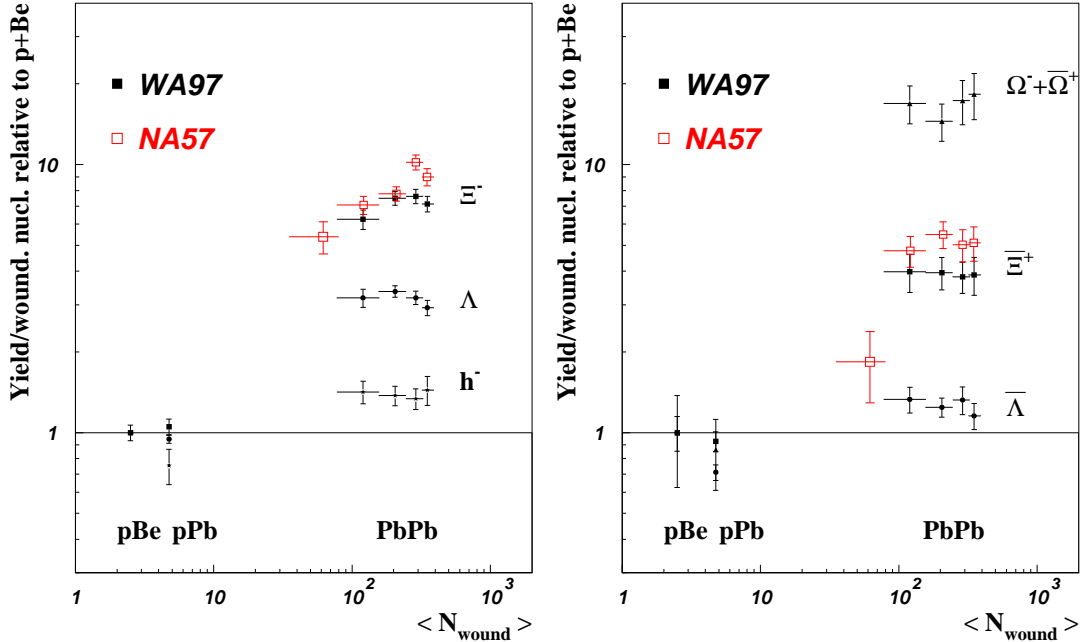}
\caption{Yields per wounded nucleons relative to p-Be
from WA97 for all measured particles (closed symbols) 
and from NA57 for $\Xi^-$ and $\overline\Xi^+$
(open symbols).\label{fig:casyie}} 
\end{figure}

\noindent The transverse mass distributions
have been parametrized according to:

\begin{equation}
\frac{d^2N}{dm_T dy}=A \hspace{1mm} m_T \exp\left(-\frac{m_T}{T}\right)
\label{eq:mtfit}
\end{equation}

\noindent and the inverse slopes $T$ have been extracted using
a maximum likelihood fit. The values are 
found~\cite{carrerqm01} to be
compatible with the WA97 ones.
No significant variation with 
the collision 
centrality is observed within
the present statistics.
\\
The particle yields have been computed by integrating
Eq.\ref{eq:mtfit} over one unit of rapidity and
extrapolating to p$_T$ = 0.
In Fig.\ref{fig:casyie} the NA57 yields for 
$\Xi^-$ and $\overline\Xi^+$
relative to the p-Be yields are shown in the five
classes, as a function of the centrality of the collision.

\noindent 
All the WA97 yields in the four most
central classes are also reported.
In the common centrality region the NA57 yields
are about 20\% larger than the WA97 ones: this
systematic difference is under investigation.
In the new low-centrality bin the 
$\overline\Xi^+$ ($\overline\Xi^-$) yields,
as measured by NA57, 
drop by a factor 2.6 (1.3) corresponding
to a 3.5 (1.8) sigma effect.
Such a sudden reduction of the yields
as a function of centrality cannot be an 
artifact of our acceptance correction procedure
since a similar drop is already present in the
uncorrected data~\cite{carrerqm01}.

\section{Summary and outlook}

Results from NA57 on cascade production
in Pb-Pb collisions at 160 $A$ GeV/$c$ 
have been reported. 
The $\Xi^-$ and $\overline\Xi^+$ yields per
participant in the most peripheral 
bin are lower
than the corresponding values in more
central collisions. The sudden drop
for $\overline\Xi^+$, in particular, 
warrants further investigation since it
could signal the point of the QGP
phase transition.
\\
The study of $\Lambda$ and $\overline\Lambda$
on the same sample is under way.
A doubling of the statisitcs with
year 2000 data will allow later this year
to perform the
analysis of the triply-strange $\Omega$ hyperons 
and to reduce the error bars on the $\Xi$ yields.
\\
The analysis of the 40 $A$ GeV/$c$ Pb-Pb data
is also under way; 
the collection of the
p-Be reference sample, needed to compute
the enhancements, will be completed in the
next summer.

\end{document}